\documentclass[reqno]{amsart}

\usepackage{epsfig}

\theoremstyle{definition}

\theoremstyle{remark}

\newcommand{\C}{\mathbb C}
\newcommand{\R}{\mathbb R}
\newcommand{\Z}{\mathbb Z}

\newcommand{\T}{\mathbb T}

\newcommand{\m}{\omega}

\newcommand{\be}{\begin{equation}}
\newcommand{\ee}{\end{equation}}
\newcommand{\lab}[1]{\label{#1}}

\begin{document}

\title[Modified elliptic gamma functions and $6d$ superconformal indices]
{Modified elliptic gamma functions \\ and $6d$ superconformal indices}

\author{Vyacheslav P. Spiridonov}
\address{Bogoliubov Laboratory of Theoretical Physics, JINR,
 Dubna, Moscow reg. 141980, Russia and Max-Planck-Institut f\"ur Mathematik,
 Vivatsgasse 7, 53111, Bonn, Germany}


\begin{abstract}
We construct a modified double elliptic gamma function
which is well defined when one of the base parameters
lies on the unit circle.
A model consisting of $6d$ hypermultiplets coupled to a
gauge field theory living on a $4d$ defect
is proposed whose superconformal index uses the double elliptic
gamma function and obeys $W(E_7)$-group symmetry.
\end{abstract}


\maketitle


\section{Introduction}

Six dimensional superconformal field theories currently form an active
research field (see, e.g., \cite{Moore} and references therein).
As claimed by Moore \cite{Moore}, these theories should form a gold mine
for experts in special functions as a source of amazing identities,
which is just one of many important potential mathematical outputs
from them. This statement sounds curious and the author agrees with it.
Indeed, a principally new class
of special functions called elliptic hypergeometric
integrals has been discovered in \cite{spi:umn}. It came as a big
surprise to mathematicians since it was tacitly assumed that
$q$-hypergeometric functions form the top level special functions
of hypergeometric type with nice exact formulas \cite{aar}.
Some particular examples of such integrals were interpreted
as wave functions or normalizations of wave functions in
specific elliptic multiparticle quantum mechanical systems \cite{spi:umn}.
Recently it was shown by Dolan and Osborn \cite{DO} that certain elliptic
hypergeometric integrals coincide with superconformal indices of four-dimensional
gauge field theories and corresponding identities prove Seiberg
dualities (electro-magnetic, strong-weak, or mirror symmetry dualities)
in the topological sector.  Further detailed investigation
of this relationship was performed in many papers
among which we mention only a small fraction \cite{SV1,SV2,GRR,DG}.

The theory of elliptic hypergeometric functions is nowadays a rich
mathematical subject with many beautiful new constructions
\cite{spi:essays}. One of its key ingredients is the
elliptic gamma function related to the Barnes multiple gamma function of
order three $\Gamma_3(u;\omega_1,\omega_2,\omega_3)$ \cite{bar}
(see the Appendix for a definition of $\Gamma_m(u;\omega)$).
The plain and $q$-hypergeometric functions
are directly related to the Barnes gamma functions of order one
$\Gamma_1(u;\omega_1)$, which is proportional to the standard
Euler's gamma function $\Gamma(u/\omega_1)$,
and order two $\Gamma_2(u;\omega_1,\omega_2)$, respectively. The multiple
infinite $q$-products with several bases naturally emerge in
the considerations of superconformal indices for higher dimensional
theories \cite{B,N}. In particular, the double elliptic gamma function
related to $\Gamma_4(u;\mathbf{\omega})$ describes
topological strings partition function
\cite{N,Iqbal,Vafa} and $6d$ superconformal indices \cite{KK,Imamura,KKK2}
(the latter was noticed also by F.~Dolan, G.~Vartanov and the author
in an unpublished consideration).
In \cite{spi:conm} the triple elliptic gamma function
(related to $\Gamma_5(u;\mathbf{\omega})$) emerged in the description
of partition functions of solvable $2d$ statistical mechanics models
identical with superconformal indices of some $4d$ quiver gauge theories.

Shortly after the discovery of elliptic beta integrals the
author posed a natural question: is there a higher order
generalization of elliptic hypergeometric integrals to
Barnes multiple gamma functions $\Gamma_m(u;\mathbf{\omega})$ with $m>3$,
obeying exact formulas similar to the ones found in \cite{spi:umn} ?
Until now this question has not been resolved, though
the author believes it has a positive answer. Perhaps
$6d$ superconformal theories provide an appropriate framework
for approaching this problem through the systematic investigation
of identities for corresponding indices.

In this note we discuss
double elliptic gamma functions and related $6d$ superconformal indices.
A modified double elliptic gamma function is introduced for which
one of the base parameters can lie on the unit circle.
The $W(E_7)$-symmetry transformation for an elliptic hypergeometric integral
established in \cite{spi:umn} and \cite{Rains} is written in a novel form
and its analogue involving the modified double elliptic gamma function is derived.
We speculate also on the structure of a theory of $\mathcal{N}=(1,0)$ $6d$
hypermultiplets coupled to an $\mathcal{N}=1$ $4d$ defect (or a similar
coupled $5d/3d$ theory) with the exact $W(E_7)$-invariant superconformal
index (or the partition function). This consideration is inspired
by analogous $4d/2d$ coupled systems of \cite{GMN} and
a $5d/4d$ boundary theory with $W(E_7)$-invariant index of \cite{DG}.

\section{Modified elliptic gamma functions}

Let us take four incommensurate quasiperiods $\omega_k\in\C$
(i.e., they are constrained by the condition $\sum_{k=1}^4n_k\omega_k\neq 0,\,
n_k\in\Z$). Using their ratios we form six bases
\begin{eqnarray}\nonumber &&
q=e^{2\pi \textup{i} \frac{\omega_1}{\omega_2}}, \quad
p=e^{2\pi \textup{i}\frac{\omega_3}{\omega_2}},\quad
r=e^{2\pi \textup{i}\frac{\omega_3}{\omega_1}},\quad
\\ &&
s=e^{2\pi \textup{i}\frac{\omega_4}{\omega_1}},\quad
t=e^{2\pi \textup{i}\frac{\omega_4}{\omega_2}}, \quad
w=e^{2\pi \textup{i}\frac{\omega_3}{\omega_4}}
\label{bases}\end{eqnarray}
and corresponding particular modular transforms
\begin{eqnarray} \nonumber  &&
\tilde q =e^{-2\pi \textup{i} \frac{\omega_2}{\omega_1}}, \quad \tilde
p=e^{-2\pi \textup{i} \frac{\omega_2}{\omega_3}},\quad \tilde r=e^{-2\pi
\textup{i}\frac{\omega_1}{\omega_3}}, \quad
\\  &&
\tilde s=e^{-2\pi \textup{i}\frac{\omega_1}{\omega_4}},\quad
\tilde t=e^{-2\pi \textup{i}\frac{\omega_2}{\omega_4}}, \quad
\tilde w=e^{-2\pi \textup{i}\frac{\omega_4}{\omega_3}}.
\label{modbases}\end{eqnarray}
The bases $p, q, r$ and  $\tilde p,\tilde q,\tilde r$ coincide
with those used in  \cite{spi:umn,spi:essays}.

In the increasing order of complexity we define the following infinite products
all of which are well defined only when bases $q,\ldots, w$ are of modulus less
than 1. Denote
$$
(z;q_1,\ldots,q_m) =\prod_{k_1,\ldots,k_m=0}^\infty(1-zq_1^{k_1}\cdots q_m^{k_m}),\ \ \ \quad z\in\C,
$$
the standard infinite $q$-product and
$
\theta(z;p)=(z;p)  (pz^{-1};p),
$
a theta function obeying properties $\theta(pz;p)=\theta(z^{-1};p)=-z^{-1}\theta(z;p)$.
The standard (order one) elliptic gamma function has the form
$$
\Gamma(z;p,q)= \prod_{i,j=0}^\infty
\frac{1-z^{-1}p^{i+1}q^{j+1}}{1-zp^iq^j}, \quad \quad z\in\C^*,
$$
and the double (i.e., of the order two) elliptic gamma function is
$$
\Gamma(z;p,q,t)= \prod_{i,j,k=0}^\infty
(1-z^{-1}p^{i+1}q^{j+1}t^{k+1})(1-zp^iq^jt^k), \quad \quad z\in\C^*.
$$
We use the conventions
\begin{eqnarray*}
&& \Gamma(a,b;\ldots):=\Gamma(a;\ldots)\Gamma(b;\ldots), \ \ \
\Gamma(az^{\pm1};\ldots):=\Gamma(az;\ldots)\Gamma(az^{-1};\ldots),
\\ &&
\Gamma(az^{\pm1}y^{\pm1};\ldots):=\Gamma(azy;\ldots)\Gamma(az^{-1}y;\ldots)
\Gamma(azy^{-1};\ldots)\Gamma(az^{-1}y^{-1};\ldots).
\end{eqnarray*}

Both, $\Gamma(z;p,q)$ and $\Gamma(z;p,q,t)$ are symmetric in their bases.
For the standard elliptic gamma function one has the difference equations
$$
\Gamma(qz;p,q)=\theta(z;p)\Gamma(z;p,q), \qquad
\Gamma(pz;p,q)=\theta(z;q)\Gamma(z;p,q).
$$
For the second order function $\Gamma(z;p,q,t)$ one has
$$
\frac{\Gamma(qz;p,q,t)}{\Gamma(z;p,q,t)}=\Gamma(z;p,t), \quad
\frac{\Gamma(pz;p,q,t)}{\Gamma(z;p,q,t)}=\Gamma(z;q,t), \quad
\frac{\Gamma(tz;p,q,t)}{\Gamma(z;p,q,t)}=\Gamma(z;p,q).
$$
The inversion relations have the form
$$
\Gamma(z,pq z^{-1};p,q)=1,\qquad  \Gamma(pqtz;p,q,t)=\Gamma(z^{-1};p,q,t).
$$

In \cite{spi:umn} the following modified elliptic gamma function was defined
\begin{eqnarray} &&
G(u;\omega_1,\omega_2,\omega_3):=\Gamma(e^{2\pi \textup{i} \frac{u}{\omega_2}};p,q)
\Gamma(re^{-2\pi \textup{i} \frac{u}{\omega_1}};r,\tilde q)
=\frac{\Gamma(e^{2\pi \textup{i} \frac{u}{\omega_2}};p,q)}
{\Gamma(\tilde q e^{2\pi \textup{i} \frac{u}{\omega_1}};r,\tilde q)}.
\label{ell-d}\end{eqnarray}
It satisfies three linear difference equations of the first order
\begin{eqnarray}\label{keyeq1}
&& G(u+\omega_1;\mathbf{\omega})=\theta(e^{2\pi
\textup{i}\frac{u}{\omega_2}};p) G(u;\mathbf{\omega}),
\\ \label{keyeq2}
&& G(u+\omega_2;\mathbf{\omega})=\theta(e^{2\pi
\textup{i}\frac{u}{\omega_1}};r) G(u;\mathbf{\omega}),
 \\ \label{keyeq3}
&& G(u+\omega_3;\mathbf{\omega})= e^{-\pi \textup{i} B_{2,2}(u;\mathbf{\omega})}
 G(u;\mathbf{\omega}),
\end{eqnarray}
where $B_{2,2}(u;\mathbf{\omega})$ is the diagonal Bernoulli polynomial of order two,
$$
B_{2,2}(u;\mathbf{\omega})=\frac{u^2}{\omega_1\omega_2}
-\frac{u}{\omega_1}-\frac{u}{\omega_2}+
\frac{\omega_1}{6\omega_2}+\frac{\omega_2}{6\omega_1}+\frac{1}{2}.
$$
In \eqref{keyeq3} the exponential coefficient emerged through
the following well-known $SL(2,\Z)$-modular transformation property
of theta functions
\begin{equation}
\theta\left(e^{-2\pi\textup{i}\frac{u}{\omega_1}};
e^{-2\pi\textup{i}\frac{\omega_2}{\omega_1}}\right)
=e^{\pi\textup{i}B_{2,2}(u;\mathbf{\omega})}
\theta\left(e^{2\pi\textup{i}\frac{u}{\omega_2}};
e^{2\pi\textup{i}\frac{\omega_1}{\omega_2}}\right).
\label{mod-theta}\end{equation}

One has the reflection equation
$
G(a;{\bf \omega})G(\omega_1+\omega_2+\omega_3-a;{\bf \omega})=1.
$
We shall use below the following shorthand notation
$$
G(a\pm b;\mathbf{\omega}):=G(a+b,a-b;\mathbf{\omega})
:=G(a+b;\mathbf{\omega})G(a-b;\mathbf{\omega}).
$$

To prove that function \eqref{ell-d} is well defined for $|q|=1$
we consider another function
\begin{equation}
\tilde G(u;\omega_1,\omega_2,\omega_3)
=e^{-\frac{\pi \textup{i} }{3} B_{3,3}(u;\mathbf{\omega})}\Gamma(e^{-2\pi
\textup{i} \frac{u}{\omega_3}}; \tilde r, \tilde p),
\label{mod2}\end{equation}
where $B_{3,3}(u;\mathbf{\omega})$ is the diagonal Bernoulli
polynomial of order three,
$$
B_{3,3}\left(u+\sum_{n=1}^3\frac{\omega_n}{2};\mathbf{\omega}\right)
=\frac{u(u^2-\frac{1}{4}\sum_{k=1}^3\omega_k^2)}{\omega_1\omega_2\omega_3}.
$$
Obviously, one has the symmetry
$\tilde G(u;\omega_1,\omega_2,\omega_3)=\tilde G(u;\omega_2,\omega_1,\omega_3).$
Using the relation
$$
B_{3,3}(u+\omega_3;\omega_1,\omega_2,\omega_3)
-B_{3,3}(u;\omega_1,\omega_2,\omega_3)=3\omega_3B_{2,2}(u;\omega_1,\omega_2),
$$
it is not difficult to check that $\tilde G(u;\mathbf{\omega})$ satisfies the same
three equations \eqref{keyeq1}-\eqref{keyeq3}
and the normalization condition
$$
\tilde G(\frac{1}{2}\sum_{k=1}^3\omega_k;\omega_1,\omega_2,\omega_3)
=G(\frac{1}{2}\sum_{k=1}^3\omega_k;\omega_1,\omega_2,\omega_3)=1.
$$
Therefore by the Jacobi theorem one obtains the equality
\begin{equation}
\tilde G(u;\omega_1,\omega_2,\omega_3)=G(u;\omega_1,\omega_2,\omega_3)
\label{tG=G}\end{equation}
corresponding to one of the $SL(3,\Z)$-modular
group transformation laws for the elliptic gamma function \cite{fel-var:elliptic}.

The crucial property of $G(u;{\bf \omega})$ is that it remains a well defined
meromorphic function of $u$ even for $\omega_{1}/\omega_2>0$
(i.e., when $|q|=1$ with the conditions $|p|,|r|<1$ being obligatory),
in difference from $\Gamma(z;p,q)$. This fact is evident from the second
form of representation of $G(u;\mathbf{\omega})$ \eqref{mod2}.

Take the limit $\omega_3\to \infty$ in such a way that
$\text{Im}(\omega_{3}/\omega_1), \, \text{Im}(\omega_{3}/\omega_2)\to +\infty$
(i.e., $p,r\to 0$). Then,
\begin{equation}
\lim_{p,r\to 0} G(u;\omega_1,\omega_2,\omega_3)
=\gamma(u;\omega_1,\omega_2)
= \frac{(e^{2\pi \textup{i} u/\omega_1}\tilde q; \tilde q) }
{(e^{2\pi \textup{i}  u/\omega_2}; q)}.
\label{2d-sin}\end{equation}
This is a modified $q$-gamma function known under many other different
names (double sine, hyperbolic gamma function, or quantum dilogarithm,
see Appendix A in \cite{spi:conm} for interconnections between these functions).
For $\text{Re}(\omega_1), \text{Re}(\omega_2)>0$ and
$0<\text{Re}(u)<\text{Re}(\omega_1+\omega_2)$ it has the integral representation
\be
\gamma(u;\omega_1,\omega_2)=
\exp\left(-\int_{\R+\textup{i} 0}\frac{e^{ux}}
{(1-e^{\omega_1 x})(1-e^{\omega_2 x})}\frac{dx}{x}\right),
\label{mod-q-gamma}\ee
which shows that $\gamma(u;\omega_1,\omega_2)$ is meromorphic
even for $\omega_1/\omega_2>0$, when $|q|=1$ and the infinite product representation
\eqref{2d-sin} is not applicable.

Euler's gamma function $\Gamma(u)$ can be defined as a special solution of
the functional equation $f(u+1)=uf(u)$. $q$-Gamma functions with
$q=e^{2\pi\textup{i}\omega_1/\omega_2}$ can be defined as special solutions
of the equation $f(u+\omega_1)=(1-e^{2\pi\textup{i}u/\omega_2})f(u)$
(in particular the functions \eqref{mod-q-gamma} and
$1/(e^{2\pi \textup{i} u/\omega_2};q)$
satisfy this equation). Analogously, the elliptic gamma functions of order
one are defined as special solutions of the key equation \eqref{keyeq1},
which does not assume any restriction on the parameter $q$. Its
particular solutions $\Gamma(e^{2\pi \textup{i} u/\omega_2};p,q)$
and $1/\Gamma(q^{-1}e^{2\pi \textup{i} u/\omega_2};p,q^{-1})$ exist only for $|q|<1$
or $|q|>1$, respectively. And $G(u;\mathbf{\omega})$ covers the remaining
domain $|q|=1$.

Define now the modified double elliptic gamma function
\begin{eqnarray}
G(u;\omega_1,\ldots,\omega_4):=
\frac{\Gamma(e^{2\pi \textup{i} \frac{u}{\omega_2}};q,p,t)}
{\Gamma(\tilde q e^{2\pi \textup{i} \frac{u}{\omega_1}};\tilde q,r, s)}.
\label{modgamma2}\end{eqnarray}
This is a meromorphic function of $u\in\C$ satisfying the inversion relation
$$
G(u+\textstyle{\sum_{k=1}^4\omega_k};\omega_1,\ldots,\omega_4)
=G(-u;\omega_1,\ldots,\omega_4)
$$
and four linear difference equations of the first order
\begin{eqnarray}\label{keyeq1mod}
&& G(u+\omega_1;\mathbf{\omega})=\Gamma(e^{2\pi
\textup{i}\frac{u}{\omega_2}};p,t) G(u;\mathbf{\omega}),
\\ \label{keyeq2mod}
&& G(u+\omega_2;\mathbf{\omega})=\Gamma(e^{2\pi
\textup{i}\frac{u}{\omega_1}}; r,s)G(u;\mathbf{\omega}),
 \\ \label{keyeq3mod}
&& G(u+\omega_3;\mathbf{\omega})=
\frac{\Gamma(e^{2\pi \textup{i} \frac{u}{\omega_2}};q,t)}
{\Gamma( \tilde q e^{2\pi \textup{i} \frac{u}{\omega_1}};\tilde q, s)}
 G(u;\mathbf{\omega}),
 \\ \label{keyeq4mod} &&
G(u+\omega_4;\mathbf{\omega})=
\frac{\Gamma(e^{2\pi \textup{i} \frac{u}{\omega_2} };p,q)}
{\Gamma(\tilde q e^{2\pi \textup{i} \frac{u}{\omega_1} };\tilde q, r)}
G(u;\mathbf{\omega}).
\end{eqnarray}
Note that the latter equation coefficient is simply $G(u;\omega_1,\omega_2,\omega_3)$.
Note also that in the limit $\omega_4\to\infty$ taken in such a way that $s,t\to 0$,
we have
$$
\lim_{s,t\to 0} G(u;\omega_1,\ldots,\omega_4)
=\prod_{j,k=0}^\infty \frac{1-e^{2\pi \textup{i} \frac{u}{\omega_2}}p^jq^k}
{1-e^{2\pi \textup{i} \frac{u}{\omega_1}}r^j\tilde q^{k+1} },
$$
which is only ``a half" of $1/G(u;\omega_1,\omega_2,\omega_3)$.

Let us demonstrate that $G(u;\omega_1,\ldots,\omega_4)$
remains a meromorphic function of $u$ for $\omega_1/\omega_2>0$
(when $|q|=1$). First, we find another
solution of the above set of equations. Consider the following function
\begin{eqnarray}
\tilde G(u;\omega_1,\ldots,\omega_4)=
e^{-\frac{\pi\textup{i}}{12}B_{4,4}(u;\mathbf{\omega})}
\frac{\Gamma(e^{-2\pi \textup{i} \frac{u}{\omega_3}};\tilde p,\tilde r,\tilde w)}
{\Gamma(w e^{-2\pi \textup{i} \frac{u}{\omega_4}};\tilde s,\tilde t, w)},
\label{modgamma2'}\end{eqnarray}
where $B_{4,4}(u;\mathbf{\omega})$ is the diagonal multiple Bernoulli polynomial
of order four, whose compact form we have found from \eqref{ber} as
\begin{eqnarray}\nonumber &&
B_{4,4}(u;\omega_1,\ldots,\omega_4)=
\frac{1}{\omega_1\omega_2\omega_3\omega_4} \Big([ (u-{\frac{1}{2}} \sum_{k=1}^4\omega_k)^2
-\frac{1}{4}\sum_{k=1}^4\omega_k^2]^2
\\ && \makebox[6em]{}
-\frac{1}{30}\sum_{k=1}^4\omega_k^4-\frac{1}{12}
\sum_{1\leq j<k\leq 4}\omega_j^2\omega_k^2\Big).
\label {B44}\end{eqnarray}
This function satisfies four linear difference equations of the first order
\begin{eqnarray}\label{keyeq1mod'} &&
\tilde G(u+\omega_1;\mathbf{\omega})=
e^{-\frac{\pi \textup{i} }{3} B_{3,3}(u;\omega_2,\omega_3,\omega_4)}
\frac{\Gamma(e^{-2\pi\textup{i}\frac{u}{\omega_3}};\tilde p,\tilde w)}
{\Gamma(w e^{-2\pi\textup{i}\frac{u}{\omega_4}};\tilde t, w)}
 \tilde G(u;\mathbf{\omega}),
\\  &&  \label{keyeq2mod'}
\tilde G(u+\omega_2;\mathbf{\omega})=
e^{-\frac{\pi \textup{i} }{3} B_{3,3}(u;\omega_1,\omega_3,\omega_4)}
\frac{\Gamma(e^{-2\pi\textup{i}\frac{u}{\omega_3}};\tilde r,\tilde w)}
{\Gamma(w e^{-2\pi\textup{i}\frac{u}{\omega_4}};\tilde s, w)}
\tilde G(u;\mathbf{\omega}),
 \\  && \label{keyeq3mod'}
\tilde G(u+\omega_3;\mathbf{\omega})=
e^{-\frac{\pi \textup{i} }{3} B_{3,3}(u;\omega_1,\omega_2,\omega_4)}
\Gamma(e^{-2\pi\textup{i}\frac{u}{\omega_4}};\tilde s,\tilde t)
 \tilde G(u;\mathbf{\omega}),
 \\  &&  \label{keyeq4mod'}
\tilde G(u+\omega_4;\mathbf{\omega})=
e^{-\frac{\pi \textup{i} }{3} B_{3,3}(u;\omega_1,\omega_2,\omega_3)}
\Gamma(e^{-2\pi\textup{i}\frac{u}{\omega_3}};\tilde p,\tilde r)
 \tilde G(u;\mathbf{\omega}),
\end{eqnarray}
following from the previously defined formulas and the
relation $B_{4,4}(u+\omega_4;\mathbf{\omega})
-B_{4,4}(u;\mathbf{\omega})=4\omega_4B_{3,3}(u;\mathbf{\omega}).$
But this is precisely the set of equations \eqref{keyeq1mod}-\eqref{keyeq4mod}.
Indeed, equality of coefficients in \eqref{keyeq4mod} and \eqref{keyeq4mod'}
is nothing else than the relation \eqref{tG=G}. Equality of
coefficients in \eqref{keyeq1mod} and \eqref{keyeq1mod'}
or in \eqref{keyeq3mod} and \eqref{keyeq3mod'}
 follows from \eqref{tG=G} after the replacement $\omega_1\to \omega_4$
or $\omega_3\to \omega_4$, respectively.
Equality of coefficients in \eqref{keyeq2mod} and \eqref{keyeq2mod'}
follows after the replacement in \eqref{tG=G} $\omega_1\to \omega_4$ and subsequent
substitution $\omega_2\to \omega_1$.
Since $\omega_j$'s are incommensurate we conclude that the ratio
$\tilde G(u;\mathbf{\omega})/G(u;\mathbf{\omega})$ is a constant independent on $u$.
However, there is no distinguished value of $u$ for which the equality of
normalizations of $G$ and $\tilde G$ becomes obvious. The fact that
\begin{equation}
\tilde G(u;\omega_1,\omega_2,\omega_3,\omega_4)
=G(u;\omega_1,\omega_2,\omega_3,\omega_4)
\label{SL4}\end{equation}
follows from an $SL(4,\Z)$-modular group transformation law
for the double elliptic gamma function established
as Corollary 9 in \cite{nar}.

So, in the same way as in the lower order cases, special solutions of
the key equation \eqref{keyeq1mod} define double elliptic gamma functions:
the functions $\Gamma(e^{2\pi\textup{i}\frac{u}{\omega_2}};p,q,t)$
and $1/\Gamma(q^{-1}e^{2\pi\textup{i}\frac{u}{\omega_2}};p,q^{-1},t)$
satisfy it for $|q|<1$ and $|q|>1$, respectively, and
$G(u;\omega_1,\ldots,\omega_4)$ covers the domain $|q|=1$.
The latter function is defined for $|p|, |r|, |s|, |t|, |w|<1$ and $|q|\leq 1$
(more precisely, for the union of the upper half plane Im$(\omega_1/\omega_2)>0$
and the half line $\omega_1/\omega_2>0$), for other admissible domains
of values of bases it will take a different form.

\section{A $6d/4d$ theory with $W(E_7)$-invariant superconformal index}

Superconformal indices are defined as \cite{KMMR,R}
$$
I(\underline{y})=\text{Tr}\, [ (-1)^F \prod_{k=1}^m y_k^{G_k}e^{-\beta H}],
$$
where $F$ is the fermion number, $G_k$ form the maximal Cartan subalgebra
preserving a distinguished supersymmetry relation involving one supercharge and
its superconformal partner
$$
\{ Q,S\}=2H, \qquad Q^2=S^2=0, \quad [Q,G_k]=[S,G_k]=0.
$$
The trace is effectively taken over the space of BPS states formed
by zero modes of the operator $H$ which eliminates dependence on
the chemical potential $\beta$. Computing supersymmetric indices
of nonconformal theories on curved backgrounds that flow to certain
superconformal field theories one gets superconformal indices of
the theories with the same superconformal fixed points
 \cite{FS}. Computation of such indices via the localization
techniques was initiated in \cite{MNS}.

We shall not discuss general structure of these indices in $4d$ field theories
since they were described in many previous papers, see, e.g.,
\cite{SV2,GRR}. Take a particular  $\mathcal{N}=1$  $4d$ theory in
the space-time $S^3\times S^1$
with $SP(2N)$ gauge group and the flavor group $SU(8) \times U(1).$
In addition to the vector superfield in
the adjoint representation of $SP(2N)$, take 8 chiral matter fields
forming the fundamental representation of $SP(2N)$ with the $R$-charge 1/2
and $U(1)$-charge $(1-N)/4$.
Take also one antisymmetric $SP(2N)$-tensor field
of zero $R$- and $SU(8)$-charges and unit $U(1)$-charge. For $N=1$, the global
group $U(1)$ decouples and the tensor field is absent.

The superconformal index of this theory is described by the following
elliptic hypergeometric integral \cite{SV1}:
\begin{eqnarray}\nonumber     && \makebox[-2em]{}
I(y_1,\ldots,y_8;t;p,q) = \frac{(p;p)^N (q;q)^N }{2^N N!}
\Gamma(t;p,q)^{N-1} \int_{{\mathbb T}^N}\prod_{1 \leq j < k
\leq N} \frac{\Gamma(t z_j^{\pm 1} z_k^{\pm 1};p,q)}
{\Gamma(z_j^{\pm 1} z_k^{\pm 1};p,q) }
    \\     &&
     \makebox[6em]{} \times
\prod_{j=1}^N \frac{\prod_{i=1}^8
\Gamma(t^{\frac{1-N}{4}}(pq)^{\frac{1}{4}} y_i z_j^{\pm
1};p,q)} {\Gamma(z_j^{\pm 2};p,q)} \frac{d z_j}{2 \pi \textup{i} z_j}.
\label{SP2N1}\end{eqnarray}
Here $y_i$ are fugacities for $SU(8)$-group satisfying the constraint
$\prod_{i=1}^8y_i=1$, $t$ is the fugacity for the group $U(1)$,
$p$ and $q$ are fugacities for the superconformal group generator
combinations $R/2 +J_1\pm J_2$, where $R$ is the $R$-charge
and $J_{1,2}$ are Cartan generators of $SO(4)$-rotations.
Nontrivial contributions to the index come only from the
states with $H=E-2J_1-3R/2=0$, where $E$ is the energy.
In terms of the variables $t_i=t^{\frac{1-N}{4}}(pq)^{\frac{1}{4}}y_i$
we have the balancing condition $t^{2N-2}\prod_{i=1}^8 t_i  = (pq)^2$.
The constraints $|t|, |t_i|<1$ are needed for the choice of the
integration contours as unit circles with positive orientation $\T$.
For $N=1$ the integral $I(y_1,\ldots,y_8;p,q)$ is nothing else than
an elliptic analogue of the Euler-Gauss hypergeometric function
introduced in \cite{spi:umn}.

In addition to the obvious $S_8$-symmetry in variables $y_i$, function
\eqref{SP2N1} obeys the following hidden symmetry  transformation
extending $S_8$-group to $W(E_7)$ -- the Weyl group of the exceptional
root system $E_7$:
\begin{eqnarray}  \label{MagnDual} \nonumber &&
    I(y_1,\ldots,y_8;t;p,q) =
\prod_{m=0}^{N-1} \Big( \prod_{1 \leq i < j \leq 4 }
\Gamma(t^{m+\frac{1-N}{2}}(pq)^{\frac{1}{2}}y_i y_j;p,q)
    \\  \nonumber   &&  \makebox[6em]{} \times
\prod_{5 \leq i < j \leq 8}
\Gamma(t^{m+\frac{1-N}{2}}(pq)^{\frac{1}{2}}y_i y_j;p,q) \Big)
I(\hat y_1,\ldots,\hat y_8;t;p,q),
\label{IM1}\end{eqnarray}
where
$$
\hat y_k =\frac{y_k}{\sqrt{Y}}, \qquad
\hat y_{k+4} =\sqrt{Y}y_{k+4}, \quad
k=1,\ldots, 4,  \qquad Y=y_1y_2y_3y_4.
$$
Equivalently one can write $Y^{-1}=y_5y_6y_7y_8$.
For $N=1$ this relation was established by the author \cite{spi:umn} and it
was extended to arbitrary $N$ by Rains \cite{Rains}.

Consider the following ratio involving double elliptic gamma functions
\begin{equation}
I_{6d/4d}(y_1,\ldots,y_8;t;p,q):=
\frac{I(y_1,\ldots,y_8;t;p,q)}
{\prod_{1\leq j<k\leq 8}\Gamma(t^{\frac{N+1}{2}}(pq)^{\frac{1}{2}}y_jy_k;p,q,t)}.
\label{6d4dind}\end{equation}
First, we show that this function is $W(E_7)$-group invariant.
Indeed, explicit substitution yields
\begin{equation}
I_{6d/4d}(\hat y_1,\ldots,\hat y_8;t;p,q)=I_{6d/4d}(y_1,\ldots,y_8;t;p,q),
\label{E7inv}\end{equation}
which follows from the relation
\begin{eqnarray*} &&
\prod_{1\leq j<k\leq 8}\frac{\Gamma(t^{\frac{N+1}{2}}(pq)^{\frac{1}{2}}y_jy_k;p,q,t)}
{\Gamma(t^{\frac{N+1}{2}}(pq)^{\frac{1}{2}}\hat y_j\hat y_k;p,q,t)}
=\prod_{1\leq j<k\leq 4 \atop 5\leq j<k\leq 8}
\frac{\Gamma(t^{\frac{N+1}{2}}(pq)^{\frac{1}{2}}y_jy_k;p,q,t)}
{\Gamma(t^{\frac{-N+1}{2}}(pq)^{\frac{1}{2}}  y_j y_k;p,q,t)}
\\  && \makebox[6em]{}
=\prod_{m=0}^{N-1}\prod_{1\leq j<k\leq 4 \atop 5\leq j<k\leq 8}
\Gamma(t^{m+\frac{1-N}{2}}(pq)^{\frac{1}{2}}y_jy_k;p,q) .
\end{eqnarray*}
In \cite{Rains} the $W(E_7)$-transformation was also written in the form
\eqref{E7inv}, but for a function different from \eqref{6d4dind}.

Now we would like to interpret equality \eqref{E7inv} as a symmetry
of the superconformal index of some $6d$ field theory with a $4d$
defect (we use the terminology of \cite{GMN} where similar mixed
$4d/2d$ theories were constructed). The main inspiration for
that comes from a beautiful $5d/4d$ field theory interpretation
of the $W(E_7)$-symmetry of the elliptic analogue of
Euler-Gauss hypergeometric function given by Dimofte and Gaiotto in \cite{DG}.

The $6d$-index for $\mathcal{N}=(1,0)$ theories on the $S^5\times S^1$
manifold is
$$
I(\underline{y};p,q,t)=\text{Tr}\, [(-1)^F p^{C_1}q^{C_2}t^{C_3}\prod_k y_k^{G_k}],
$$
where $G_k$ are the flavor group maximal torus generators and $C_{1,2,3}$
are Cartan generators for the space-time symmetry group.
In the notations of Imamura \cite{Imamura}
$$
p^{C_1}q^{C_2}t^{C_3}=x^{j_1+3R/2}y_3^{j_2}y_8^{j_3},
$$
where $R$ is the Cartan of $SU(2)_R$-subalgebra, $j_1$ is the generator
of $U(1)_V$ and $j_2, j_3$ are Cartans of $SU(3)_V$ with $U(1)_V\times SU(3)_V$
being a subgroup of the $SO(6)$-isometry group of $S^5$.
Perturbative contributions to the index are described by the double elliptic
gamma functions \cite{Imamura,KKK2} with bases
$$
p=\frac{xy_3}{y_8},\quad q=\frac{x}{y_3y_8},\quad t=xy_8^2.
$$
One can permute $p, q,$ and $t$, but we stick to this choice
leading to
\begin{equation}
C_{1,2}=\frac{1}{3}(j_1-\frac{j_3}{2})\pm \frac{j_2}{2}+\frac{R}{2}, \qquad
C_3=\frac{1}{3}(j_1+j_3)+\frac{R}{2}.
\label{Cops}\end{equation}
E.g.,  for a $U(1)$-flavor group hypermultiplet one has the index
\begin{eqnarray}\label{hyper} &&
I_{hyp}(y;p,q,t)=\frac{1}{\Gamma(\sqrt{pqt}y;p,q,t)}
=\exp\Big(\sum_{n=1}^\infty\frac{i_{hyp}(y^n;p^n,q^n,t^n)}{n}\Big), \quad
\\  &&
i_{hyp}(y;p,q,t)=\frac{\sqrt{pqt}(y+y^{-1})}{(1-p)(1-q)(1-t)}.
\nonumber \end{eqnarray}
For $SU(2)$ gauge group vector superfield one obtains%
\begin{eqnarray}\label{vector} &&
I_{vec}(z;p,q,t)=\kappa\frac{\Gamma(z^{\pm 2};p,q,t)}{(1-z^2)(1-z^{-2})}
=\exp\Big(\sum_{n=1}^\infty\frac{i_{vec}(z^n;p^n,q^n,t^n)}{n}\Big), \quad
\\ \nonumber &&
i_{vec}(z;p,q,t)=\Big(1-\frac{1+pqt}{(1-p)(1-q)(1-t)}\Big)
\chi_{adj,\, SU(2)}(z),
\\  &&
\kappa=\lim_{x\to 1}\frac{\Gamma(x;p,q,t)}{1-x}=(p;p)(q;q)
(t;t)(pq;p,q)(pt;p,t)(qt;q,t)(pqt;p,q,t)^2.
\nonumber \end{eqnarray}
The multiplier $\kappa$ appears naturally from the adjoint representation
character $\chi_{adj,\, SU(2)}(z)=z^2+z^{-2}+1$.
One can incorporate into $I_{vec}$ a piece of Haar measure for $SU(2)$
and cancel thus the terms $(1-z^2)(1-z^{-2})$.

Take now the $4d$ interacting gauge theory described above and
assume that it lives on a $S^3\times S^1$ manifold immersed
into the taken $6d$ space-time $S^5\times S^1$
in the ``corner" $x_5=x_6=0$ of $S^5$ defined by the coordinate constraint
$\sum_{i=1}^6x_i^2=1$.
This defect breaks half of $6d$ supersymmetries, presumably preserving the
supercharge used for defining the superconformal index. Associate fugacities
$p$ and $q$ with the isometries of the space $S^3\times S^1$, which
connects corresponding $4d/6d$ Cartan generators as $J_1\propto j_1-j_3/2$
and $J_2\propto j_2$. The abelian group associated with the generator
$C_3$ in \eqref{Cops} is identified from the $4d$ theory point of view
with the $U(1)$-flavor group whose fugacity is $t$.
Take now free $6d$ gauge invariant  hypermultiplets forming the
totally antisymmetric tensor of second rank $T_A$ for the mentioned
$SU(8)$ flavor group and couple them to the taken $4d$ defect.

The interaction superpotential that ties together the flavor symmetries and the
rotation symmetry of the defect could be of the form
$W = {\rm Tr}\, q M q\big|_{x_5=x_6=0}$,
where $q$ are the $4d$ quark superfields and $M$ is
one of the two chiral fields in the $6d$ hypermultiplet.
Here the trace contracts the gauge indices of the quarks with the symplectic form
as well as the $SU(8)$ flavor indices of the quarks and $M$.
Since $q$ has $U(1)$-charge $(1-N)/4$, this results in
the additional $U(1)$-charge of the hypermultiplets equal to $N/2$
after taking into account the rotation symmetries of $M$ and the actual
Lagrangian couplings obtained after integration of $W$ over the superspace
(the author is indebted to D. Gaiotto for pointing out on such a possibility).
This yields a $6d$ model with a $4d$ defect similar to $4d/2d$
systems considered in \cite{GMN} (in particular, see the toy model
considered in Sect. 4.3 of \cite{GGS}).
As a result, the hypermultiplet index takes the form
\begin{equation}
I_{T_A}(\underline{y};p,q,t)=\prod_{1\leq j< k\leq 8}
\frac{1}{\Gamma(t^{N/2}\sqrt{pqt}y_jy_k;p,q,t)}, \quad \prod_{k=1}^8y_k=1,
\label{indTA}\end{equation}
which evidently coincides with the multiplier in \eqref{6d4dind}.
The ``corner" defect $4d$ theory gives its own contribution
to the superconformal index described by the integral
$I(y_1,\ldots,y_8;t;p,q)$.
The combined index has $W(E_7)$-symmetry
indicating that this theory may have the enhanced $E_7$-flavor
group, provided there exists an appropriate point in the moduli space.
This is a rough potential physical picture behind
relation \eqref{E7inv} the detailed consideration of which lies
beyond the scope of the present note.
For $N=1$ a simplification takes place since the $U(1)$ group decouples
from the $4d$-defect. In this case $I_{6d/4d}$
turns into the $W(E_7)$-invariant half-index of \cite{DG} in the limit
$t\to 0$, but this seems to be a formal coincidence since the
$t$-parameter should stay intact in the half-index for $N>1$.

As proposed in \cite{spi:umn}, one can build elliptic hypergeometric integrals
using the modified elliptic gamma function. This is achieved by mere
replacement of $\Gamma(e^{2\pi\textup{i}u/\omega_2};p,q)$-functions by
$G(u;\omega_1,\omega_2,\omega_3)$ and appropriate change of the
integration contour. Taking the limit $\omega_3\to \infty$ such that $p,r\to 0$
one obtains hyperbolic hypergeometric integrals expressed in terms
of the hyperbolic gamma function $\gamma^{(2)}(u;\omega_1,\omega_2)$
(see the Appendix).
Using this procedure the integral \eqref{SP2N1} can be reduced \cite{ds:unit,DSV}
to the following expression:
\begin{eqnarray}\nonumber     && \makebox[-2em]{}
I_h(x_1,\ldots,x_8;\lambda;\omega_1,\omega_2) = \frac{1}{2^N N!}
\gamma^{(2)}(\lambda ;\omega_1,\omega_2)^{N-1}
\\ \nonumber && \makebox[-2.5em]{}
\times \int_{-\textup{i} \infty}^{\textup{i} \infty} \prod_{1 \leq j <
k \leq N} \frac{\gamma^{(2)}(\lambda \pm u_j \pm
u_k;\omega_1,\omega_2)}{\gamma^{(2)}(\pm u_j \pm
u_k;\omega_1,\omega_2)} \prod_{j=1}^N \frac{\prod_{k=1}^{8}
\gamma^{(2)}(\mu_k \pm u_j;\omega_1,\omega_2)}{\gamma^{(2)}(\pm 2
u_j;\omega_1,\omega_2)} \prod_{j=1}^{N} \frac{d u_j}{\textup{i}
\sqrt{\omega_1 \omega_2}},
\label{SP2Nhyper}\end{eqnarray}
where chemical potentials are related to flavor fugacities
as $t=e^{2\pi\textup{i}\lambda/\omega_2}$ and $y_k=e^{2\pi\textup{i}x_k/\omega_2}$ with
$$
\mu_k= x_k +\frac{\omega_1 + \omega_2}{4}- (N-1)\frac{\lambda}{4},
\qquad \sum_{k=1}^8 x_k=0.
$$
In terms of $\mu_k$ the balancing condition reads
$$
2(N-1)\lambda + \sum_{k=1}^{8} \mu_k  = 2 (\omega_1 + \omega_2).
$$
Define now a new function
\begin{equation}
I_{5d/3d}(x_1,\ldots,x_8;\lambda;\omega_1,\omega_2)
=\frac{I_h(x_1,\ldots,x_8;\lambda;\omega_1,\omega_2)}
{\prod_{1\leq j<k\leq 8}\gamma^{(3)}(\frac{N+1}{2}\lambda+\mu_j+\mu_k;
\omega_1,\omega_2,\lambda)},
\label{5dPF}\end{equation}
where $\gamma^{(3)}(u;\omega_1,\omega_2,\lambda)$ is the hyperbolic gamma function
of third order (see the Appendix).
Again, it is not difficult to see that this function is $W(E_7)$-invariant
as a consequence of known relations for $I_h$-integral:
\begin{equation}
I_{5d/3d}(\hat x_1,\ldots,\hat x_8;\lambda;\omega_1,\omega_2)
=I_{5d/3d}(x_1,\ldots,x_8;\lambda;\omega_1,\omega_2),
\label{5d_ident}\end{equation}
where
$$
\hat x_k=x_k -\frac{1}{2}\sum_{l=1}^4x_l, \qquad
\hat x_{k+4}=x_{k+4} +\frac{1}{2}\sum_{l=1}^4x_l, \quad k=1,\ldots, 4.
$$

Integral \eqref{5dPF} may be interpreted as the partition
function of some $5d$-theory coupled to a $3d$ defect. Indeed,
contribution of $5d$ hypermultiplets to the partition function is determined by the
$1/\gamma^{(3)}$-function \cite{Vafa,Imamura} which indicates that our
$5d$ theory has the field content similar to the one described earlier with
the defect $S^3\times S^1$ replaced by the squashed three-sphere
$S^3_b$ with $b^2=\omega_1/\omega_2$. The transition from $4d$ indices
to $3d$ partition functions of theories living on such manifolds was
described in \cite{DSV}. Note that symmetry transformation
\eqref{5d_ident} looks similar to the enhanced $E_n$-global
symmetry discussed in \cite{KKL} asking for an investigation of
potential relations between corresponding theories.

\section{Relevance of the modular group transformations}

Let us discuss physical relevance of modular groups
acting on the generalized gamma functions. Quasiperiods $\omega_k$ are usually
interpreted as squashing parameters and coupling constants.
The generalized gamma functions are defined differently
for different domains of these parameters related to each other by modular
transformations usually playing the role of $S$-dualities.

The simplest example of the relevance of $SL(2,\Z)$-modular group is given
by the $q$-gamma function. It can be defined as a solution of the equation
$f(u+\omega_1)=(1-e^{2\pi\textup{i}u/\omega_2})f(u)$.
For $|q|<1$ its solution $1/(e^{2\pi\textup{i}u/\omega_2};q)$
defines the standard $q$-gamma function and serves as a building block
of various partition functions. However, to cover the region $|q|=1$,
one needs $SL(2,\Z)$-modular transformation  \cite{fad} and define
the modified $q$-gamma function \eqref{2d-sin},
i.e. to use the ratio of modular transformed elementary partition functions.

Consider now the elliptic gamma function $\Gamma(z;p,q)$
describing the superconformal index for a $4d$ chiral superfield.
In order to define an analogue of this function for the region $|q|=1$
in \cite{spi:umn} the modified elliptic gamma function was proposed
as the ratio of this index with a $U(1)$-group fugacity
parametrization $z=e^{2\pi\textup{i} u/ \omega_2}$ and superconformal
group generator fugacities $q=e^{2\pi\textup{i} \omega_1/ \omega_2}$
and $p=e^{2\pi\textup{i} \omega_3/ \omega_2}$ and the index with a
different choice of squashing parameters
$\Gamma(\tilde q e^{2\pi \textup{i} \frac{u}{\omega_1}};r,\tilde q)$.
Surprisingly, this ratio yields again the chiral field index
with yet another parametrization of fugacities
$e^{-\frac{\pi \textup{i} }{3} B_{3,3}(u;\mathbf{\omega})}\Gamma(e^{-2\pi
\textup{i} \frac{u}{\omega_3}}; \tilde r, \tilde p)$. The
exponential cocycle factor spoils this interpretation and requires
a physical interpretation. As shown in \cite{SV2} this $SL(3,\Z)$-group
action on $4d$ superconformal indices describes
the 't Hooft anomaly matching conditions as the conditions of cancellation
of this cocycle contributions described by a curious set of Diophantine equations.
Therefore this modular group plays quite important role in the formalism.

A similar picture at the level of free $6d$ hypermultiplet index
was described recently in \cite{Vafa} in relation to the topological
strings partition function. Namely, $I_{hyp}(y;p,q,t)$ is proportional to
the latter function and, as argued in \cite{Vafa}, a
particular combination of three $SL(4,\Z)$-transformed versions of it
should yield yet another similar partition function. And this expectation
is confirmed with the help of an $SL(4,\Z)$-modular group
transformation for the double elliptic gamma function which
is written in our case as equality \eqref{SL4}.

However, in difference from the $G(u;\omega_1,\omega_2,\omega_3)$-function
case, the elliptic hypergeometric integrals formed from
$G(u;\omega_1,\ldots,\omega_4)$ do not reduce to the integrals
composed from $\Gamma(z;p,q,t)$. Now the modular
group simply maps them into similar integrals up to the cocycle
$\propto e^{-\frac{\pi\textup{i}}{12}B_{4,4}}$ multiplying the
integral kernels. Therefore one should not expect cancellation
of these factors from the integrals. Cancellation of
even the gauge group chemical potentials is
possible only under very strong restrictions, e.g. for $SU(2)$ gauge group
it is possible only for $N_f=16$ at the expense of an unusual
quadratic restriction on chemical potentials. Such exponentials
have the forms resembling the Casimir energy contributions to the
indices \cite{KKK2}. Therefore it is necessary to better understand the
general structure of full $6d$ superconformal indices before
connecting $SL(4,\Z)$-modular group transformations to higher
dimensional anomalies. Still, we can see an involvement of the
$B_{4,4}$-polynomial in the $4d$ anomaly matching conditions.

Define a modified elliptic hypergeometric integral:
\begin{eqnarray}\nonumber     && \makebox[-2em]{}
I^{mod}(x_1,\ldots,x_8;\omega_1,\ldots,\omega_4) =
\frac{(\tilde p;\tilde p)^N (\tilde r;\tilde r)^N }{2^N N!}
G(\omega_4;\omega_1,\omega_2,\omega_3)^{N-1}
    \\  \nonumber   &&      \makebox[4em]{} \times
\int_{[-\frac{\omega_3}{2},\frac{\omega_3}{2}]^N}\prod_{1 \leq j < k \leq N}
\frac{G(\omega_4\pm u_j\pm u_k;\omega_1,\omega_2,\omega_3)}
{G(\pm u_j\pm u_k;\omega_1,\omega_2,\omega_3) }
    \\     &&      \makebox[4em]{} \times
\prod_{j=1}^N \frac{\prod_{i=1}^8
G(x_i -\frac{N}{4}\omega_4 +\frac{1}{4}\sum_{k=1}^4\omega_k\pm u_j;
\omega_1,\omega_2,\omega_3)} {G(\pm2 u_j;\omega_1,\omega_2,\omega_3)}
 \frac{d u_j}{\omega_3},
\label{SP2Nmo}\end{eqnarray}
which is obtained from \eqref{SP2N1} simply by the replacement of
$\Gamma$-functions by $G$-functions using exponential representation
for fugacities in terms of chemical potentials and passing to the
integration over a cube. Note that this integral is well-defined for $|q|=1$.
Introduce ``the modified index"
\begin{eqnarray}\nonumber &&
I_{6d/4d}^{mod}(x_1,\ldots,x_8;\omega_1,\ldots,\omega_4)
   \\ &&      \makebox[4em]{}
= \frac{I^{mod}(x_1,\ldots,x_8;\omega_1,\ldots,\omega_4)}
{\prod_{1\leq j<k\leq 8}G(\frac{N}{2}\omega_4+\frac{1}{2}\sum_{k=1}^4\omega_k+x_j+x_k;
\omega_1,\ldots,\omega_4)},
\label{6d4dindmo}\end{eqnarray}
containing the modified double elliptic gamma function.
It is not difficult to check that this expression is also $W(E_7)$-invariant
\begin{equation}
I_{6d/4d}^{mod}(\hat x_1,\ldots,\hat x_8;\omega_1,\ldots,\omega_4)
=I_{6d/4d}^{mod}(x_1,\ldots,x_8;\omega_1,\ldots,\omega_4).
\label{E7invmo}\end{equation}
Now one can replace $G$-functions by their modular transformed expressions
$\tilde G$ containing exponentials of Bernoulli polynomials
and check that relation \eqref{E7invmo} boils down to
an $SL(3,\Z)$-modular transformation of the previous relation \eqref{E7inv}.
At the level of integral \eqref{SP2Nmo} with the constraint
$2(x_7+x_8)=\sum_{k=1}^3\omega_k+(N-1)\omega_4$ this was done already
in \cite{ds:unit}. As mentioned above,
the condition of cancellation of Bernoulli polynomial coefficients
in the integration variables and external parameters
describes the 't Hooft anomaly matching conditions. Therefore the
fourth order polynomial $B_{4,4}(u;\mathbf{\omega})$ is effectively
involved into these anomaly matchings as well.

The residue calculus for elliptic hypergeometric integrals
was developed long ago, see \cite{spi:umn,spi:essays} and
references therein. It has shown that by shrinking the
integration contour to zero one can formally represent integrals
as sums of bilinear combinations of elliptic hypergeometric series
with permuted base variables which describes the factorization of
superconformal indices into some more elementary building blocks
which in general are not defined in the limit $p\to 0$ or $q\to 0$.
This analysis has lead to the discovery of the
notion of two-index biorthogonality and the elliptic modular
doubling principle \cite{spi:umn,spi:essays}.
In \cite{GRR} this residue calculus applied to $4d$ $\mathcal{N}=2$
superconformal indices was physically interpreted as a result of insertions
of surface defects into the bulk theory.

One can investigate the structure of residues for the modified
elliptic hypergeometric integrals/indices and come to similar
factorization in terms of different elliptic hypergeometric
series. The latter series are related by an $SL(3,\Z$)-transformation
and remain well defined in the
limit $p\to 0$, which leads to hyperbolic integrals.
As a result hyperbolic integrals are represented
as combinations of products of two $q$-hypergeometric series
related by an $SL(2,\Z)$-modular transformation (their
bases are $q$ and $\tilde q$) \cite{spi:umn,spi:essays}. This
factorization was used in \cite{SVlmp} for computing partition functions
in some $3d$ $\mathcal{N}=2$ theories appearing from the reduction of $4d$
$\mathcal{N}=4$ SYM theories. The principle difference between
$4d$ (elliptic) and $3d$ (hyperbolic) cases consists in the fact
that in $3d$ this factorization of sums of residues into modular
blocks has rigorous meaning because of the convergence of
corresponding infinite series
for $|q|<1$, whereas in $4d$ such series do not converge
for generic values of $p$ and $q$ bases and the factorization
of indices has in general a formal meaning.
It is not difficult to develop the residue calculus
for $6d$ indices and find triple sums of residues.
However, corresponding sums cannot factorize because there are no
triply periodic functions.
This makes the $4d$ (elliptic) case rather unique and raises
the interest to $6d$ indices as qualitatively different objects.

\smallskip

The author is indebted to T. Dimofte, D. Gaiotto, Y. Imamura, A. Klemm, J. Manschot,
G. W. Moore, and G. S. Vartanov for valuable discussions and to the referee for
constructive remarks.
This work is partially supported by RFBR grant
no. 11-01-00980 and NRU HSE scientific fund grant no. 12-09-0064.

\appendix
\section{Barnes multiple gamma function}

Barnes multiple zeta function $\zeta_m(s,u;\mathbf{\omega})$ \cite{bar}
is originally defined by an $m$-fold series
$$
\zeta_m(s,u;\mathbf{\omega})=\sum_{n_1,\ldots,n_m=0}^\infty
\frac{1}{(u+\Omega)^s}, \qquad\Omega=n_1\omega_1+\ldots+n_m\omega_m,
$$
where $s, u\in\C$. It converges for $\text{Re}(s)>m$, provided
all $\m_j$ lie in one half-plane formed by a line passing through zero
(then there are no accumulation points of the $\Omega$-lattice
in compact domains).

This zeta function satisfies equations
\be
\zeta_m(s,u+\omega_j;\mathbf{\omega})-\zeta_m(s,u;\mathbf{\omega})
=-\zeta_{m-1}(s,u;\mathbf{\omega}(j)), \quad j=1,\ldots,m,
\lab{zeta-eq}\ee
where $\mathbf{\omega}(j)=(\omega_1,\ldots,\omega_{j-1},\omega_{j+1},\ldots,
\omega_m)$ and $\zeta_0(s,u;\mathbf{\omega})=u^{-s}$.
The multiple gamma function is defined by Barnes as
$$
\Gamma_m(u;\mathbf{\omega})
=\exp(\partial \zeta_m(s,u;\mathbf{\omega})/\partial s)\big|_{s=0}.
$$
As a consequence of \eqref{zeta-eq} it satisfies $m$ finite difference equations
\begin{equation}
\Gamma_m(u+\omega_j;\mathbf{\omega})
=\frac{1}{\Gamma_{m-1}(u;\mathbf{\omega}(j))}\, \Gamma_m(u;\mathbf{\omega}),
\qquad j=1,\ldots,m,
\label{bar-eq}\end{equation}
where  $\Gamma_0(u;\omega):=u^{-1}$.

The multiple sine-function is defined as
$$
S_m(u;\mathbf{\omega})=\frac{\Gamma_m(\sum_{k=1}^m\m_k-u;\mathbf{\omega})^{(-1)^m}}
{\Gamma_m( u;\mathbf{\omega})}
$$
and the hyperbolic gamma function is
$$
\gamma^{(m)}(u;\mathbf{\omega})=S_m(u;\mathbf{\omega})^{(-1)^{m-1}}.
$$
One has equations
$$
\gamma^{(m)}(u+\omega_j;\mathbf{\omega})
= \gamma^{(m-1)}(u;\mathbf{\omega}(j))\, \gamma^{(m)}(u;\mathbf{\omega}),
\qquad j=1,\ldots,m.
$$
The standard elliptic gamma function can be written as
a special ratio of four $\Gamma_3(u;\mathbf{\omega})$-functions,
and the double elliptic gamma function is given by
a product of four $\Gamma_4(u;\mathbf{\omega})$-functions \cite{spi:essays}.

One can derive the integral representation \cite{nar}
\begin{eqnarray*} &&
\gamma^{(m)}(u;\mathbf{\omega})=
\exp\left( -\text{PV}\int_{\R} \frac{e^{ux}}
{\prod_{k=1}^m(e^{\omega_k x}-1)}\frac{dx}{x}\right)
\\ && \makebox[2em]{}
= \exp\left(- \frac{\pi \textup{i}}{m!}B_{m,m}(u;\mathbf{\omega})
-\int_{\R+\textup{i}0}\frac{e^{ux}}
{\prod_{k=1}^m(e^{\omega_k x}-1)}\frac{dx}{x}\right)
\\ && \makebox[2em]{}
= \exp\left(\frac{\pi \textup{i}}{m!}B_{m,m}(u;\mathbf{\omega})
-\int_{\R-\textup{i}0}\frac{e^{ux}}
{\prod_{k=1}^m(e^{\omega_k x}-1)}\frac{dx}{x}\right),
\end{eqnarray*}
where Re$(\omega_k)>0$ and $0<\text{Re}(u)< \text{Re}(\sum_{k=1}^m\omega_k)$
and $B_{m,m}$ are multiple Bernoulli polynomials defined by the generating function
\be
\frac{x^m e^{xu}}{\prod_{k=1}^m(e^{\m_k x}-1)}
=\sum_{n=0}^\infty B_{m,n}(u;\m_1,\ldots,\m_m)\frac{x^n}{n!}.
\lab{ber}\ee
In particular, one has the following relation with the modified
$q$-gamma function $\gamma(u;\omega_1,\omega_2)$:
$$
\gamma^{(2)}(u;\omega_1,\omega_2)=e^{-\frac{\pi \textup{i}}{2}B_{2,2}(u;\omega_1,\omega_2)}
\gamma(u;\omega_1,\omega_2).
$$
Collapsing integrals to sums of residues one can derive infinite product
representations for $\gamma^{(m)}(u;\mathbf{\omega})$ \cite{nar}.
Particular inversion relations have the form
\begin{eqnarray*} &&
\gamma^{(2)}(\textstyle{\sum_{k=1}^2\omega_k}+u;\omega_1,\omega_2)
\gamma^{(2)}(-u;\omega_1,\omega_2)=1, \quad
\\ &&
\gamma^{(3)}(\textstyle{\sum_{k=1}^3\omega_k}+u;\omega_1,\omega_2,\omega_3)=
\gamma^{(3)}(-u;\omega_1,\omega_2,\omega_3).
\end{eqnarray*}

\end{document}